\documentclass[conference]{IEEEtran}
\usepackage{xcolor}
\usepackage{multirow}
\usepackage{graphicx} 
\usepackage{listings}
\usepackage{algorithm}
\usepackage{algorithmicx}
\usepackage{algpseudocode}
\ifCLASSINFOpdf
\else
\fi

\begin{document}
%
\title{Hunting the Ghost: Towards Automatic Mining of IoT Hidden Services}

\newcommand\dsk[1]{\{\textbf{dsk:}\textcolor{purple}{\em#1}\}}

\newcommand\cjy[1]{\{\textbf{cjy:}\textcolor{blue}{\em#1}\}}

\newcommand{\system}{\texttt{IoTBolt}\space}

\author{\IEEEauthorblockN{Shuaike Dong}
\IEEEauthorblockA{The Chinese University of Hong Kong\\
1155085891@link.cuhk.edu.hk}
\and
\IEEEauthorblockN{Siyu Shen}
\IEEEauthorblockA{The Chinese University of Hong Kong\\
ss019@ie.cuhk.edu.hk}
\and
\IEEEauthorblockN{Zhou Li}
\IEEEauthorblockA{University of California, Irvine\\
zhou.li@uci.edu}
\and
\IEEEauthorblockN{Kehuan Zhang}
\IEEEauthorblockA{The Chinese University of Hong Kong\\
khzhang@ie.cuhk.edu.hk}}


%


\IEEEoverridecommandlockouts
\makeatletter\def\@IEEEpubidpullup{6.5\baselineskip}\makeatother
\IEEEpubid{\parbox{\columnwidth}{
    Network and Distributed Systems Security (NDSS) Symposium 2020\\
    23-26 February 2020, San Diego, CA, USA\\
    ISBN 1-891562-61-4\\
    https://dx.doi.org/10.14722/ndss.2020.23xxx\\
    www.ndss-symposium.org
}
\hspace{\columnsep}\makebox[\columnwidth]{}}

\maketitle

\begin{abstract}
The Internet of Things (IoT) has been prosperously
used and profoundly changing people’s daily life. However, for
normal users, most IoT devices are smart “black boxes” listening
to their commands and providing feedbacks. Recent notorious
IoT-based attacks have shown that even IoT devices themselves
may behave normally, evil things can be happening under the
hood. Therefore, it is crucial for both normal users and security
analysts to quickly profile the behavior of an IoT device and dig
out ”open” and ”half-open” doors in it. In this paper, we proposes
IoTBolt , an automatic firmware analysis tool targeting at
finding hidden services that may be potentially harmful to the IoT
devices. Our approach uses static analysis and symbolic execution
to search and filter services that are transparent to normal users
but explicit to experienced attackers. A prototype is built and
evaluated against a dataset of IoT firmware, and The evaluation shows our tool can find the suspicious hidden services effectively.
\end{abstract}


%

\section{Background}
\subsection{Smart-Home IoT Architecture}
A typical smart home consists of three parties: IoT device, service provider and user. The normal interaction scheme is shown in fig~\label{fig:smart_home_arch}. Below we briefly introduce it.

\begin{figure}[htb]
	\centering
	\includegraphics[width=0.8\columnwidth]{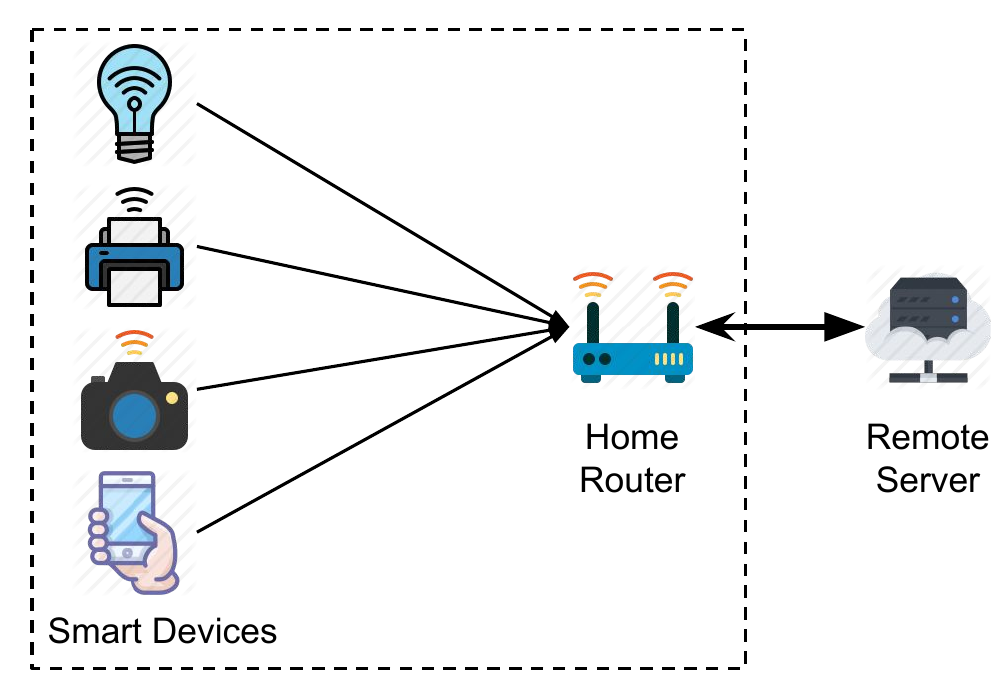}           
	\caption{A typical smart home}
	\label{fig:smart_home_arch}
\end{figure} 

\subsection{IoT Firmware}
IoT firmware is the software programmed on IoT hardware. Since IoT products are usually embedded devices with limited computation resources and variant architectures. IoT firmware provides basic functionalities to interact with the hardware. 

\noindent \textbf{Firmware layout.} There are two typical types of firmware, \textit{firmware blobs} and \textit{embedded-Linux based firmware}. Firmware blobs often come with bare-metal devices with very monotonous capabilities and restricted resources, like smart bulbs and smart plugs. Firmware blobs usually have various formats due to the proprietary development process of their vendors.

According to previous works~\cite{DBLP:conf/uss/CostinZFB14}, embedded-Linux based firmware takes up the largest portion of off-the-shelf firmware with a ratio of around 86\%. Contents of an embedded-Linux based firmware vary among different vendors. In most cases, there is at least one embedded file system inside and can be extracted by firmware unpacking tools like \textit{binwalk}~\cite{url_binwalk}. 

\noindent \textbf{Firmware acquisition.} There are generally three ways to obtain a firmware. 

1) Download from websites of the vendor. Several vendors release the firmware dataset on their websites, like \textit{D-Link}~\cite{url_dlink_repo} and \textit{TP-Link}~\cite{url_tplink_download_center}.

2) Extract firmware from hardware. Firmware often resides in the storage unit of an IoT device such as {flash chip}. With microcontrollers that supports flash communication protocols, one can dump the firmware without destroying the device~\cite{url_dump_firmware}. Apart from flash chip dumping, other hardware debugging ports can be leveraged for the firmware extraction as well~\cite{url_extract_firmware}.

3) Network interception. Some vendors provide mobile Apps to facilitate the control of IoT devices. Most of these Apps can trigger the firmware upgrade by one or several simple operations. During the firmware upgrading process, either the meta-data (including download url, file size and checksum, etc.) or the firmware itself will be transmitted to the device. An experienced analyst can easily intercept the traffic he/she needs by setting up a proxy.

\subsection{Issues in IoT Development}
As the development of a commercial IoT product is a long and lasting task during which the code may experience a lot of changes and updates from different developers. Many issues may arise during the process and thus produce some "hidden" services, which may be utilized by malicious attackers. We classify the factors to these risks as follows: 
\begin{enumerate}
    \item \textbf{Leftover Debugging code.} As the modern software development is usually a complicated process including multiple coupling components, developers widely use debugging to locate the errors and fix them. Different from codes released in production version, debugging code usually involves a wider capability and higher privilege, or some functionalities that are just for testing usage. Apart from that, the debugging codes are usually not directly visible to normal users but can be easily found by experienced and purposeful attackers.
    
    \item \textbf{Unmatched interfaces and services.} To facilitate normal users in using IoT devices, vendors usually provide easy-to-use interfaces for them. For example, most domestic routers provide a series of web pages leading users to configure the product step by step. As those pages are just interactive entrances to back-end running programs, we call them \texttt{interfaces} and their targets as \texttt{services}. From the perspective of a normal user, interfaces and services should have a "surjection" relation which means each service should have at least one interface to trigger it. However, there are cases when multiple services running behind the scene without any companion interfaces. Apart from that, even those "already-matched" services may contain hidden functions that are not within the scope of their interfaces, which are exploitable by experienced attackers.
    
    
    
    \item \textbf{Backdoors.} Previous work~\cite{DBLP:conf/ndss/Shoshitaishvili15} has shown that backdoors are pervasive in IoT devices. From the perspective of vendors, backdoors are helpful for debugging usage and testing phase during the development. However, they can be taken advantage of by attackers and thus leading to the compromise of the whole device or privacy exposure.
    
\end{enumerate}

\subsection{IoT interfaces and services}
\label{sec:iot_interfaces_and_services}
As the medium between normal users and the environment, IoT devices provide easy-to-use interfaces for consumers. We categorize common interfaces as follows:

\begin{enumerate}
	\item \textbf{Web-based interfaces.} Some IoT devices embed web pages inside the firmware to help users do the configuration. For example, home-use routers often host the domain ``http://192.168.0.1'' by a web server. People access that domain and set up the router step by step according to its wizard. According to our study, some IoT cameras, also support such kind of interface.
	\item \textbf{Mobile-based interfaces.} More and more IoT developers are releasing their companion mobile Apps with devices. A user yields commands by operating the UI controllers of the mobile App, the commands will be transferred to the cloud or directly sent to the device in the form of network packets.
	\item \textbf{Physical contacts.} Most IoT devices provide physical interfaces to make them consistent with previous non-IoT devices, like traditional plugs and sockets. Some physical actions like long-pressing a button also act as the critical role in the binding process, like ~\cite{DBLP:conf/dsn/ChenZDDZSLZZ19} shows. 
	\item \textbf{Others.} There are additional interfaces utilizing different kinds of signals. For example, Google home~\cite{url_googlehome} responds when it hears legal voice commands. Some smart thermometers commits certain behaviors when the temperature has reached a threshold.
\end{enumerate}

After a user request is well received by the device, it is processed by the software components of the IoT firmware. Due to the variety of the information involved inside requests, different handlers are chosen to react. We refer to reactions taken by the IoT devices as \texttt{services}, which are IoT functionalities that can be aroused through IoT interfaces. 

To conclude, IoT services hold the following attributes: 
\begin{enumerate}
	\item can be triggered through IoT interfaces or channels with the equal semantics. For instance, a service of a home router can be activated either by interacting with the web interface or by a manually-crafted network packet.
	\item can reveal information to the sender of a request.
	\item can change the device status.
\end{enumerate}

\noindent \textbf{Hidden Service.} Due to the ``black box" feature of an IoT device, we differentiate a hidden service from a normal one by \textit{whether it is transparent to normal users}. To simplify, we assume normal users have no enough knowledge of network security and can only operate IoT devices following specifications. For example, when given a home-use router, a normal user only performs the following steps:
\begin{enumerate}
	\item Plug in necessary wires and turn on the router.
	\item Open the browser and access ``http://192.168.0.1'', input default credentials on the specification to authorize.
	\item Follow the Setup Wizard and configure the router.
	\item Connect his/her electronic devices to the WiFi and surf the Internet.
\end{enumerate}
He or she is unaware of more advanced techniques like penetration tests. According to our definition, all the functionalities provided outside the scope of ``http://192.168.0.1'' are hidden from the user, and therefore are hidden services.

Note that hidden services do not certainly mean vulnerabilities. However, the existence of hidden services brings unnecessary redundancy to the IoT system and has negative influence on its robustness and security.
\section{System Overview}
\label{sec:system_overview}
\system\  aims at finding hidden services in IoT devices. In this section, we introduce the high-level idea of our tool and how it works in an automatic way.

\begin{figure}[htb]
	\centering
	\includegraphics[width=0.9\columnwidth]{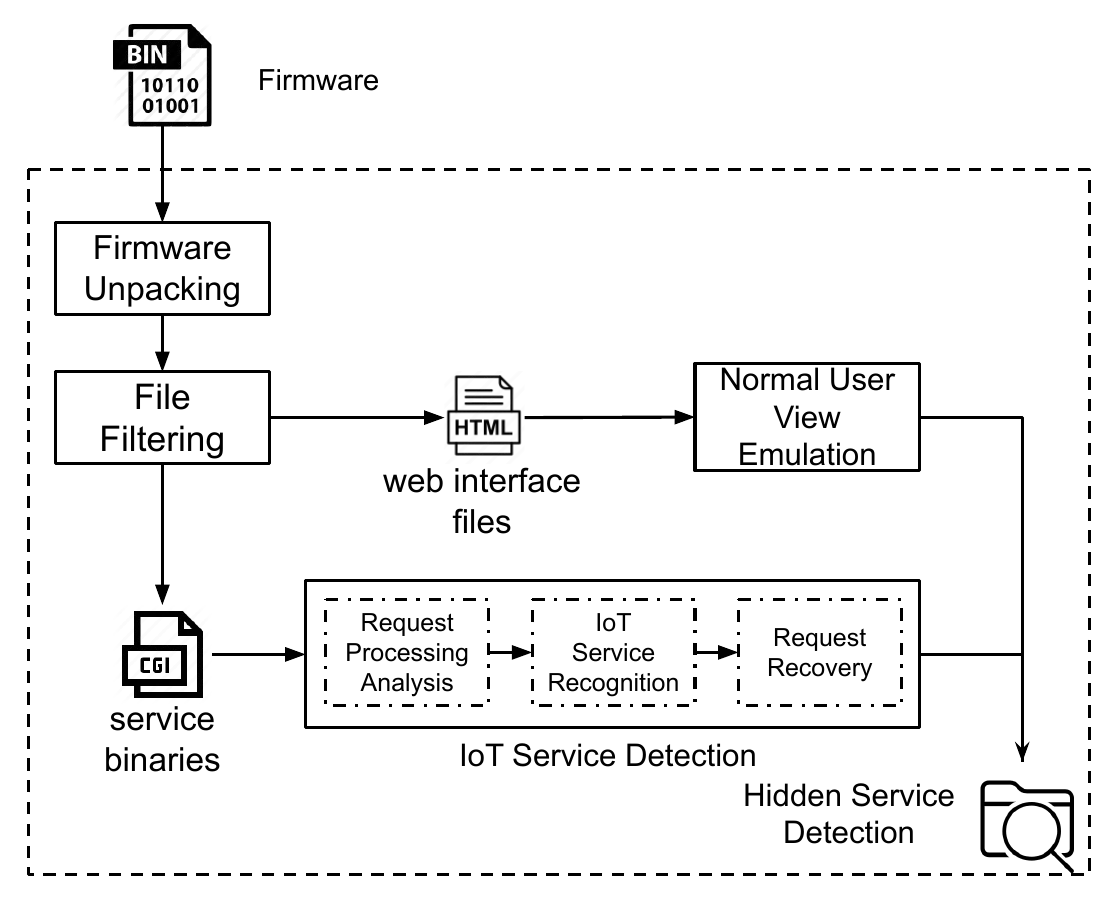}           
	\caption{The workflow of \texttt{IoTBolt}}
	\label{fig:workflow}
\end{figure}

Figure~\ref{fig:workflow} shows the overall structure of \system. From it we know, \system takes 4 steps analyzing each input firmware sample.

\begin{enumerate}
	\item \textbf{Firmware Unpacking.} \system works on Linux-based firmware. We use the popular firmware extraction tool \texttt{binwalk} to unpack the firmware. After a successful unpacking, all files in the original file system (\texttt{squash-fs}, etc.) of the firmware are restored.
	\item \textbf{Files Filtering.} Step 1 will recover most of the files inside a file system, including executables, libraries, configuration files, ect.. However, not all of them are needed by \system. \system will recognize all web-based interface files, including those with extensions like \textit{.html, .js, .php, .asp}. \system will also identify executables handling user requests and provide services, which we call \textit{service binaries}. The detailed algorithm we use is illustrated in section~\ref{sec:files_filtering}.
	
	\item \textbf{IoT Services Recognition.} After locating \textit{service binaries}, \system analyzes them in a static way to find and summarize the provided services. At this step, \system also prepares for the symbolic execution by identifying the start and end points.
	
	\item \textbf{Normal User View  Emulation.} 
	
	\item \textbf{Hidden Services Detection.}
\end{enumerate}

\section{Files Filtering}
\label{sec:files_filtering}
\system takes a straightforward name-based approach to filtering the files needed.

\noindent \textbf{Document Root Location.} Current web servers usually hold most of their web interface files in a certain directory, called \textit{Document Root}. Files under the Document Root directory are what a user sees when he/she is visiting a website. 

To recognize the document root directory, we count the number of files with extensions like \textit{.html}, \textit{.js}, \textit{.php}, \textit{.asp}, \textit{.css}. The directory holding the largest quantity of web interface files is regarded as the document root.

\noindent \textbf{Service Binary Recognition.} We categorize the \textit{service binaries} into two folds: \textit{server-like binaries} and \textit{module-like binaries}. These two types mainly differentiate at \textit{whether it can receive/send network requests/responses individually}. Except from web servers, there are some daemon programs belong to server-like binaries. They provide other interfaces, like another open port rather than 80 and 443, to interact with users.

Module-like binaries mainly refer to Common Gateway Interface (CGI) files written in C programming language. These files can interact with users but only with CGI-modules enabled by web servers.

We recognize service binaries by counting the occurrences
of a set of network-related strings. We also perform a scanning
in /etc folder, if those service binaries are found in those
initialization files, like /init.d/rcS, we assume they have larger
chances to be service binaries and prioritize them in theanalyzing queue.
\section{Request Processing Analysis}
\label{sec:requests_processing}
\begin{figure}[htb]
	\centering
	\includegraphics[width=0.9\columnwidth]{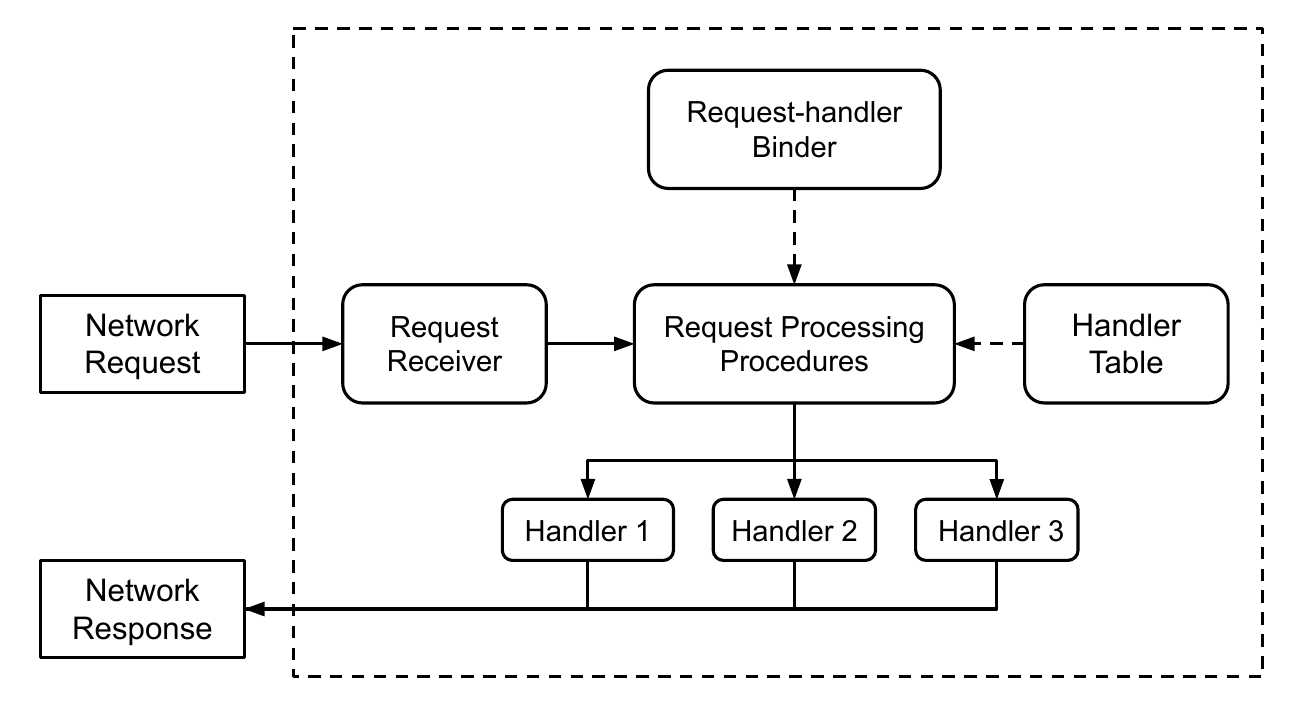}           
	\caption{The procedure of processing a network request}
	\label{fig:request_handlers}
\end{figure} 

As section~\ref{sec:iot_interfaces_and_services} says, services can be triggered only when requests received at the firmware side can be successfully processed. Figure~\ref{fig:request_handlers} shows a normal process of handling a network request. At first, the request is captured by a \textit{request receiver}, which is often realized by standard C functions like \textit{recv} and \textit{recvfrom}. After that, the request is parsed and checked by \textit{request processing procedures}.

\begin{figure}[htb]
	\centering
	\includegraphics[width=0.9\columnwidth]{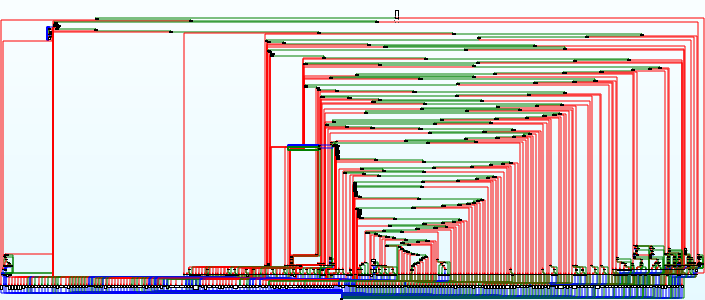}           
	\caption{The overview graph of a typical request processing procedure(by IDA)}
	\label{fig:overview_graph}
\end{figure}

A request processing procedure examines network messages by traversing a tree-like structure. At each internal node of the tree, a certain field inside the message is checked and different values will lead to different paths until a corresponding service is invoked. Figure~\ref{fig:overview_graph} shows the control flow graph of a typical request processing procedure.

Previous works have already proposed several approaches to identifying request processing procedures. For example, Karonte~\cite{redini2020karonte} leveraged five features to evaluate whether a function implements a request processing procedure or not, including (1) the number of basic blocks, (2) the number of branches, (3) the number of conditional statements used in conjunction with memory comparisons, (4) \textit{network mark}, the number of network-related strings that are used by memory comparisons and (5) \textit{connection mark}, whether the data received from a network socket is used in memory comparisons.

Karonte's approach has moved a big step forward to the automatic recognition of network-related parsing functions. However, we find there are still some limitations that are not fixed, which are listed as below:

\begin{enumerate}
	\item Some embedded web servers provide APIs to associates a URL with a specific handler.  For example, \textit{GoAhead}~\cite{url_goahead} provides an API called \textit{websUrlHandlerDefine} whose first parameter is a string representing the target URL and the second parameter is a processing function object, whenever the target URL is detected, the corresponding handler will be invoked to process the request. We refer to such functions as \textit{request-handler binders} since server itself will initiative deliver requests to their handlers without querying a tree-like structure. 
	
	\item The heavy-weight approach to calculate the \textit{connnection mark} can easily fail. According to the details of Karonte, we find it applies symbolic-execution-based taint analysis to trace the data flow from standard socket APIs, like \textit{recv} and \textit{recvfrom} to memory comparisons like \textit{memcmp} and \textit{strcmp}. This approach can work when the program being analyzed is small-sized. However, for a program with a large size and many \textit{indirect control flow transitions}, the approach can easily gets lost and stops in the half way. The following code snippet shows how \texttt{boa} hinders the taint analysis with the usage of a message queue.
\end{enumerate}

To fix the limitations above, we propose the following improvements to Karonte's approach based on our observations, which are summarized as below:

\begin{itemize}
	\item \textbf{Observation 1.} A request-handler binder usually contains a network-related string and a function object as parameters. Apart from that, request-handler binders are usually invoked for multiple times to bind different URLs with their handlers.
	\item \textbf{Strategy 1.} We propose another two features, of which the first recording whether a network-related string and a function object appear together as the arguments of a subroutine, the second being the number of invocations to this subroutine.
	
	\item \textbf{Observation 2.} For a request processing subroutine with a tree-like structure, the request or the object representing the request is usually stored in function arguments or global variables.
	 
	\item \textbf{Strategy 2.} Instead of tracking the data flow from standard socket APIs, we determine to apply intra-procedure taint analysis on the subroutine. \system checks if the parameters and the global variables appearing inside the subroutine are finally used by memory comparisons.
	
\end{itemize}

\noindent \textbf{Function table recognition.} Apart from the tree-like structure, \textit{function tables} are often used to find corresponding handlers. A function table is often composed of multiple compactly-stacked items. Each item has a key(name) and a value(handler address). Once a key matches that of a request, the corresponding handler will be invoked. Table~\ref{tab:function_table} shows a real function table we extract from our firmware database and Listing~\ref{list:algorithm} illustrates how function table is usually used inside a function.

\begin{table}[]
	\scalebox{1.15}[1.2]{
\begin{tabular}{|c|c|c|}
	\hline
	\textbf{address} & \textbf{name}                   & \textbf{handler} \\ \hline
	0x58c560         & ``SetMultipleActions" (0x4a6620) & sub\_433768      \\ \hline
	0x58c568         & ``GetDeviceSettings" (0x4a6634)  & sub\_423d28      \\ \hline
	0x58c570         & ``GetOperationMode" (0x4a6648)   & sub\_433f70      \\ \hline
	0x58c578         & ...                             & ...              \\ \hline
	0x58c888         & 0x00                            & 0x00             \\ \hline
\end{tabular}
	}
	\vspace{0.1cm}
	\caption{A typical function table in memory.}
	\label{tab:function_table}
\end{table}

\begin{algorithm}[htb]
	\begin{algorithmic}[1]
	\Require $url \gets the\ url\ to\ access$
	\State $pointer \gets \&func\_table$
	\While{$*pointer \neq zero\_byte$}
		\If {$*pointer == url$}	
			\State $handler \gets *(pointer+4)$

    			\Call{Call}{handler}
		\Else
			\State $pointer = pointer + 8$
		\EndIf
	\EndWhile
	\end{algorithmic}
\caption{Abstract of function table usage}
\label{list:algorithm}
\end{algorithm}

Locating function tables can significantly enrich the services we can find. Based on our observations, function tables are often accessed inside loop structures. To this end, we take the following steps to locate them.

1) Identify \textit{loop} structures inside a function body. This can be done by \textit{angr}'s built-in analysis modules.

2) Partially symbolic-execute the loop until an ending (usually byte zero) is encountered. Collect all memory addresses accessed during the execution and store them into a list $[addr_1, addr_2, ..., addr_n]$.

3) Generally, the collected memory addresses are those pointing to name strings, therefore, we calculate the $stride$ between $addr_i$ and $addr_{i+1}$ and get a second address list $[addr_1+stride/2, addr_2+stride/2, ..., addr_n+stride/2]$ 

4) For each address inside the second address list, we match the disassembly code at that address against the built-in function prologues of \textit{angr}. A successful match implies an identification of a function table item, and we finally obtain a function table containing multiple jumping targets.

\section{IoT Services Recognition}
\label{sec:service_recognition}
Services are the most critical part of an IoT device due to their usability. In this section, we introduce how \system recognize the services within a firmware. Note that the ``service'' in our paper emphasizes on its availability, not the concrete behaviors that can benefit people's daily lives, like ``playing a song'' and ``turning on the light''.

\subsection{Service Categorization}
Referring to attributes of services listed in section~\ref{sec:requests_processing}, we summarize the services into four categories:

\noindent \textbf{persistent output.} One of the main capabilities of IoT devices is to feed back users with information they have asked for. After manually analyzing multiple firmwares from different vendors, we conclude three strategies that are commonly used in network-related binaries.

\begin{enumerate}
	\item For Common Gateway Interface (CGI) files, information is directly output by standard output functions, like \textit{printf} and \textit{puts} in C programming language, \textit{echo} in \textit{Linux bash}.
	\item For non-CGI files, especially web server binaries, information is usually yielded to files using stream output functions
    , like \textit{fprintf} and \textit{fputs}. Whether the user can access the information is determined by the authentication and authorization settings of that binary.
	\item Some developers leverage third-party libraries to deliver data of different formats from plain HTML, like Extensible Markup Language (XML) and JavaScript Object Notation (JSON). For example, the web server file \textit{goahead} in a D-Link firmware\footnote{DLink DIR823G\_V1.0.2B05\_20181207}widely uses functions from library \textit{libmxml.so} to get/set values inside a XML object.
\end{enumerate}

\noindent \textbf{system command execution.} Since a firmware contains multiple binaries with different capabilities, there are often cases when a user's request can only be fulfilled by executing another binary. Apart from that, Linux provides a lot of easy-to-use but powerful commands, which are frequently-used in IoT firmwares, like \textit{reboot}, \textit{sleep} and \textit{echo}, etc. For example, when a user uploads a new version of firmware to the home router, the router will first store the new firmware in a temporary area and changes the bootloader settings. After that, a \textit{system("reboot")} will be executed to make the device reboot and load the newer firmware.

According to our observation, the most commonly-used system command execution APIs are \textit{system} and \textit{exec} function family, including \textit{execle}, \textit{execv}, etc.

\noindent \textbf{non-volatile storage operation.} Non-volatile storage is pervasively used in IoT devices due to its small size and low power consumption~\cite{url_nonvolatile}. As the physical carrier of firmware itself, non-volatile storage is often used for device configuration and user personalization~\cite{url_advantages_nonvolatile}. Developers can also control the non-volatile storage with the help of non-volatile memory libraries(\textit{libnvram.so}, etc.) like operating a normal key-value storage. Different from a normal key-value storage, the updates is stored globally and can be shared among different binaries.

\noindent \textbf{network activities.} We conclude two scenarios when IoT devices create network activities. (1) \textbf{Network activities between IoT devices and remote servers.} Due to the restricted resources of an IoT device, some of its functionalities can only be accomplished with the help of remote servers. For example, a smart assistant like \texttt{Google Home} and \texttt{Amazon Alexa} have to interact with remote servers to answer users' various questions. (2) \textbf{Network activities between IoT devices and terminal users.} One typical case is network cameras with real time streaming protocol(RTSP) supported . When the user accesses the RTSP channel of the camera, the camera will start sending RTSP packets continuously.

\subsection{Service Recognition}

\begin{table}[]
	\scalebox{1.0}[1.1]{
		\begin{tabular}{|c|c|c|ll}
			\cline{1-3}
			\textbf{service category}                                               & \textbf{APIs}                                                                                                & \textbf{library needed}                                                 &  &  \\ \cline{1-3}
			\multirow{4}{*}{persistent output}                                      & printf, printf\_s, puts                                                                                      & \multirow{5}{*}{C standard library}                                     &  &  \\ \cline{2-2}
			& fprintf, fprintf\_s, fputs                                                                                   &                                                                         &  &  \\ \cline{2-2}
			& fopen, fopen64                                                                                               &                                                                         &  &  \\ \cline{2-2}
			& open, open64                                                                                                 &                                                                         &  &  \\ \cline{1-2}
			\begin{tabular}[c]{@{}c@{}}system command\\ execution\end{tabular}      & \begin{tabular}[c]{@{}c@{}}system, execl, execle, execlp\\ execv, execve, execvp, popen\end{tabular}                &                                                                         &  &  \\ \cline{1-3}
			\begin{tabular}[c]{@{}c@{}}nonvolatile storage\\ operation\end{tabular} & \begin{tabular}[c]{@{}c@{}}nvram\_set, nvram\_update\\ apmib\_set, apmib\_update\end{tabular}                & libnvram.so/apmib.so                                                    &  &  \\ \cline{1-3}
			network activities                                                      & \begin{tabular}[c]{@{}c@{}}bind, connect, SSL\_connect\\ SSL\_read, SSL\_write\end{tabular} & \begin{tabular}[c]{@{}c@{}}C standard library;\\ libssl.so\end{tabular} &  &  \\ \cline{1-3}
		\end{tabular}
	}
	\vspace{0.1cm}
	\caption{Typical APIs of four IoT services}
	\label{tab:service_api}
\end{table}

We locate services by searching for APIs with correct contexts, which we call \textbf{end points}. The selected APIs are listed in table~\ref{tab:service_api}. Note the API list can be extended when some third party libraries are used in the development.

\noindent \textbf{API context.} Even for the same set of APIs, there is no guarantee that the consequences of their usage are the same. For example, although \textit{printf} can output information as HTML codes directly in CGI files, non-CGI files do not have such feature, where \textit{printf} is frequently used to yield messages to the console, for the aim of logging and debugging. So is stream output function \textit{fprintf}~\cite{url_fprintf}. Given its prototype -- \textit{int fprintf (FILE * stream, const char * format, ...);}, we know the first parameter plays a big role in determining the behavior of \textit{fprintf}. The following piece of code gives an example. From it we see, although \textit{fprintf} writes ``something'' to the file \textit{\_\_stream}, however, \textit{\_\_stream} refers to \textit{/dev/console} that does not deliver any information to the user's screen.

\system deals with this by considering and recovering the context of API call sites. For \textit{printf}-like functions, \system determines whether it is an end point by the type of the file, in our case, CGI or non-CGI. For \textit{fprintf}-like functions, \system first searches for corresponding file handling functions, like \textit{open} and \textit{fopen}, in the vicinity of \textit{fprintf} call sites. After that, \system recovers its arguments, only when the argument \textit{filename} is not the console device and the argument \textit{access mode} includes ``writing'' capability, will the invocation be regarded as a legal end point.

{\small
	\begin{lstlisting}[
	caption={\textit{fprintf} output to console},
	captionpos=b
	language={C},
	label={lst:xxx},
	keywordstyle=\color{blue!70},
	numbers=left,                    
	numbersep=5pt, 
	xleftmargin=8pt,
	xrightmargin=5pt,
	numberstyle=\scriptsize\color{gray},
	breaklines=true,
	frame=single,
	basicstyle=\ttfamily, 
	commentstyle=\color{blue} \textit,
	stringstyle=\ttfamily, 
	showstringspaces=false,
	mathescape=true]
$\color{red!70}\textbf{\_\_stream}$ = fopen("$\color{red!70}\textbf{/dev/console}$", "w");
fprintf($\color{red!70}\textbf{\_\_stream}$,'SCRIPT NAME{%s}\n', __s1); 
	\end{lstlisting} 
}

\noindent \textbf{API encapsulation.} Another issue we consider is the encapsulation of APIs. According to our observations, dynamically linked shared object libraries may introduce functions with the same capabilities but different names from the APIs listed in table~\ref{tab:service_api}. For example, we found \textit{libputil.so} of the firmware \textit{D-Link DIR-505 v1.09B02} defines a function called \textit{\_system}, which encapsulates the API \textit{system} by adding a call to \textit{vnsprintf} in the front. Two functions behave in the almost same way but are invoked interchangeably in other binaries, as shown in the listing~\ref{lst:_system}. To fix this, \system scans the whole file system(in the firmware) looking for dynamically linked shared libraries used by the binary being analyzed, if there are any function found to be the encapsulation of the APIs in table~\ref{tab:service_api} and are invoked inside the current binary, \system add them to the end points list.


{\small
	\begin{lstlisting}[
	caption={\textit{fprintf} output to console},
	captionpos=b
	language={C},
	label={lst:_system},
	keywordstyle=\color{blue!70},
	numbers=left,                    
	numbersep=5pt, 
	xleftmargin=8pt,
	xrightmargin=5pt,
	numberstyle=\scriptsize\color{gray},
	breaklines=true,
	frame=single,
	basicstyle=\ttfamily, 
	commentstyle=\color{blue} \textit,
	stringstyle=\ttfamily, 
	showstringspaces=false,
	mathescape=true]
int _system(char* param_1, char* param_2)
{
  char arr[520];
  int var = $\color{red!70}\textbf{vsprintf}$(arr, param_1, param_2);
  $\color{red!70}\textbf{system}$(arr);
  return Var;
}       
	\end{lstlisting} 
}

\section{Requests Recovery by Symbolic Execution}
\label{sec:requests_recovery}

After identifying services inside a binary, \system tries to recover the network requests that can trigger them. In this section, we describe the challenges we met in recovering the requests and how we overcome them.

\noindent \textbf{\textit{Challenge 1}. Determine start points of symbolic execution.} As a popular software testing approach, symbolic execution engine needs to specify an entry point to start the simulation. By default for a C-language program, the \textit{main} function will be set as the entry point. However, for a lot of complex binaries like web servers, \textit{main} function is not the optimal solution, since there are a lot of initialization and pre-processing tasks within the \textit{main} function which can be the burden to the symbolic execution. 

\noindent \textbf{\textit{Solution 1.}} The intuition to solve challenge 1. is that \textit{request processing subroutine} is the most important in recovering the request. If a request can trigger the service along a certain path, the \textit{request processing subroutine} must be on that path as well.

Following this, we first generate a call graph for the whole binary. For each end point collected as the section~\ref{sec:service_recognition} shows, we perform a backward tracing from it until a function with no predecessors is found. Among all functions in the trace, we select the one holding the highest probability of being the \textit{request processing subroutine}, and take it as the start point for that end point.

\noindent \textbf{\textit{Challenge 2.} State explosion during symbolic execution.} State explosion has been a long-term challenge when applying symbolic execution to practical projects~\cite{SurveySymExec-CSUR18}. For network-related binaries, the issue becomes obvious due to the extensive existence of loop-like structures and deep function call traces.

\%dsk{to think about why our work is more difficult}

\noindent \textbf{\textit{Solution 2.}} Our high-level idea of solving challenge 2. is to leverage static analysis to guide the symbolic execution. To be specific, we statically generate a set of \textit{chopped control flow graphs}  for each pair of start point and end point. The graphs are used to guide the system in avoiding unnecessary and unrelated paths during the symbolic execution. The detailed algorithm is introduced in section~\ref{sec:requests_recovery}. Apart from that, we take the strategy of \textit{high-level constraints recording}. On one hand, this strategy helps the symbolic execution engine bypass functions that have no or little impacts on request recovery, on the other hand, provide security analysts with clearer information about the request contents rather than byte- or bit-level memory constraints.

\subsection{Multi-stage Chopped Control Flow Graphs}
\label{sec:chopped}
As the aim of symbolic execution engine is to rapidly explore the paths from a start point to an end point and recover the critical information inside the request body that leads to it, strategies should be taken to guide the symbolic engine to avoid unrelated paths. To achieve this, we leverages static analysis to generate multi-stage chopped control flow graphs before running the symbolic execution.

We referenced the concept of ``chopped CFG" proposed by Brumley et al.~\cite{DBLP:conf/csfw/BrumleyWJS07}. To reduce the overhead of chopped CFG generation, we take advantage of function-level backward trace mentioned in Solution 1. To be specific, given one pair of start point and end point, instead of calculating chopped CFGs from the whole binary, we decompose the task into multiple sub-tasks. That is to say, for each caller and callee along the backward trace, we set the caller's address as the start point and the callee's call site as the end point, then calculate the sub-chopped-CFG of them. After iterating the whole trace, we compose all sub-chopped-CFGs to get the \textit{multi-stage chopped CFG}. Taking advantage of the backward trace significantly reduce the time spending on getting chopped CFGs. 

During the symbolic execution, the current symbolic state will be checked against the multi-stage chopped CFG at each step. If the basic block being accessed by the current state is inside its basic block set, we assume the state is moving towards the target end point in the right direction, otherwise, the state will be dropped for the sake of appropriate memory cost. Listing~\ref{alg:multi-stage chopped CFG} shows the detailed algorithm.

\begin{algorithm}[htb]
	\begin{algorithmic}[1]
		\Require \\
		$endpoint$: the service end point; \\
		$callgraph$: the call graph of a function; \\
		$cfg$: the control flow graph of the whole binary
		
		\Function {BackwardTrace}{$endpoint, callgraph$}
		\State get the function involving the $endpoint$
		\State $func \gets endpoint.function$
		\State $traces \gets DFS(endpoint)$ //Use Depth-First-Search to find all function traces to $endpoint$
		
		\State \Return $traces$
		\EndFunction
		
		\Function {genChoppedCFG}{$traces$}
		\State $addrs \gets []$ 
		\For {$trace$ in $traces$}
		\For {$i=0$; $i<trace.length-1$; $i++$}
		\State $chopped \gets DFS(trace_{i}, trace_{i+1})$
		\State $addrs = addrs + chopped$
		\EndFor
		\EndFor
		\State \Return $addrs$
		\EndFunction
	\end{algorithmic}
	\caption{The Algorithm of generating multi-stage chopped CFGs}
	\label{alg:multi-stage chopped CFG}
\end{algorithm}

\subsection{High-level Constraints Recording}

Due to the exploration scheme mentioned in section~\ref{sec:chopped}, \system limits its symbolic execution to codes that directly deals with request parsing and service triggering. This can not fully simulates the running of a real-world network-related binary. For instance, a web server usually does some initialization work during its start up, like reading configurations from local file system and setting up necessary environment variables. Due to such ``tedious'' jobs are usually not included inside the selected code paths for \system, the symbolic solver is prone to fail in determining the possible values when the constraints contain those unknown values.

Another consideration is the external functions and unrelated functions along the code path. On one hand, diving into these functions might significantly increase the time and memory cost of the symbolic execution, on the other hand, these functions can have influence on the further execution of a normal program.

To fix the issues above, we take the strategy of \textit{high-level constraints recording}. The strategy includes two aspects:

First, \system provides an int-type user-controlled parameter called \texttt{step-depth}. This parameter decides whether the symbolic execution engine should step into a function and by default it is set to zero for the sake of high efficiency. For a function to be bypassed , \system will assign a symbolic variable to the register containing the return value of that function. All these symbolic variables will participate in the following symbolic execution.
Using this approach, on one hand, we ignore the internal details of a function so that the chance of getting stuck in ``local'' subroutines can be significantly reduced,
on the other hand, provides a high-level abstract of data flow for the further execution by symbolizing the return value of that function. The user-controlled parameter can be tuned when needed, for example, analyzing small-sized binaries with not-too-many branches.

For function arguments, we only symbolize the arguments of the start point function. The reason comes from the observation that for a network-related binary, requests or request-like objects are more frequently ``read'' (compared with certain keywords/values) than ``written'' by a function. In that case, bypassing a certain function usually does not hinder us from recovering critical information of a request, like its keywords and URLs. 

Second, we create a tree data structure recording high-level constraints along a certain path. The motivation is that \texttt{angr}'s solver engine \texttt{Claripy}~\cite{url_claripy} and \texttt{Z3}~\cite{url_z3} back-end provide byte- or bit-level constraints, which are hard for analysts to understand~\cite{ndss:human_readable}.
For example, the code in Listing~\ref{lst:high-level} produces 31 constraints at 0x46e534, among which there are tedious descriptions of certain memory locations like Listing~\ref{lst:tedious_constraints} shows.

{\small
	\begin{lstlisting}[
	caption={An example of a tedious constraints by angr symbolic execution engine},
	captionpos=b
	language={C},
	label={lst:tedious_constraints},
	keywordstyle=\color{blue!70},
	numbers=left,                    
	numbersep=5pt, 
	xleftmargin=8pt,
	xrightmargin=5pt,
	numberstyle=\scriptsize\color{gray},
	breaklines=true,
	frame=single,
	basicstyle=\ttfamily, 
	commentstyle=\color{blue} \textit,
	stringstyle=\ttfamily, 
	showstringspaces=false,
	mathescape=true]
<Bool !(!(!(48 <= mem_fff00002_39_8{UNINITIALIZED})||(!(mem_fff00002_39_8{UNINITIALIZED}[5:0] <= 57)||!(mem_fff00002_39_8{UNINITIALIZED}[7:6] == 0)))||(!(num_bytes_38_32 == 0x1)||(!(mem_fff00001_37_8{UNINITIALIZED}[7:6] == 0)||... (mem_fff00002_39_8{UNINITIALIZED}[5:0] <= 57) ||(!(mem_fff00001_37_8{UNINITIALIZED}[5:0] <= 57)||!(48 <= mem_fff00001_37_8{UNINITIALIZED})))))))))))))))))))>
	\end{lstlisting} 
}
The constraints generated by \texttt{angr}'s symbolic execution are hard to understand by analysts. To supplement its constraint representation, we \textit{record the critical high-level semantics along a certain code path} and \textit{restore the dependencies among high-level constraints}. The approach works in three steps:

1) During the symbolic execution at each basic block, \system maintains a set of \textit{symbolic variables}. If a symbolic variable is generated by bypassing the function as previously mentioned, its arguments will also be recorded. In Listing~\ref{lst:high-level}, we show a real example from \textit{boa}\footnote{DLink DIR-619L\_B1\_FW206WWb01\_f58h}. The code has been manually recovered to C language from the disassembly.

2) 
When a branch instruction (\textit{jz} and \textit{jnz} in x86, \textit{bez} and \textit{bnez} in MIPS, etc.) is hit,
there are usually two successor states generated, one following the ``True'' branch and the other taking the ``False'' branch. We select the constraints from two states that own the same set of variables and totally-opposite operations (like ``=='' and ``!=''), recovered the arguments and record them. In the case below, they are \textit{strcmp(\_\_nptr, ``0'')==0} and \textit{strcmp(\_\_nptr, ``0'')!=0}. 

\begin{figure}[htb]
	\centering
	\includegraphics[width=0.9\columnwidth]{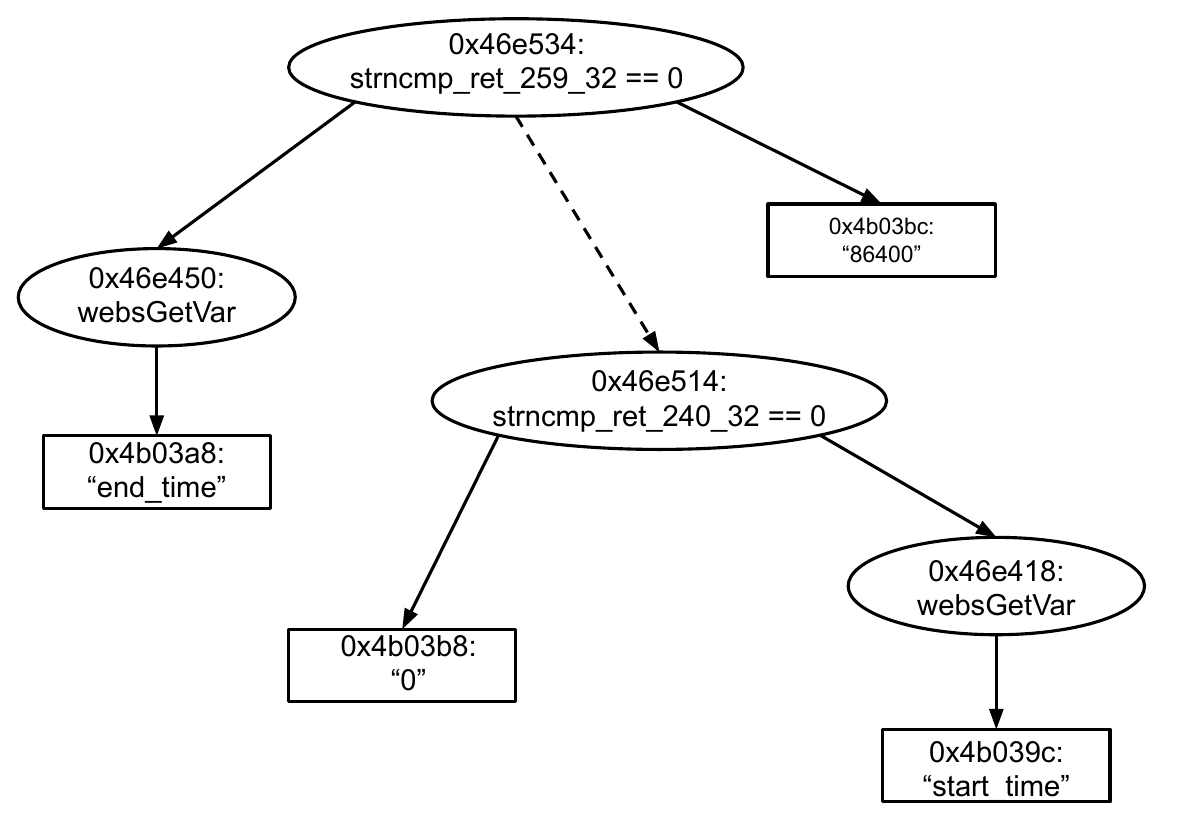}           
	\caption{The high-level constraints collected by \system}
	\label{fig:high-level}
\end{figure}

3) When the symbolic engine has reached the end point, we iterate all the basic blocks in its access history, collect all the high-level constraints and construct the constraint dependency tree. The final tree structure can clearly show the data dependency along a certain code path, which is shown in figure~\ref{fig:high-level}. 

Although \texttt{angr} does not record dependencies between constraints explicitly, we restore them by two means -- on one hand, the recovered arguments of a function call maintained an implicit connection of two constraints, shown by the edge between node ``0x46e450:websGetVar'' and node ``0x46e534:strcmp == 0'' in Figure~\ref{fig:high-level}, on the other hand, during the symbolic execution, we collect the current address of a symbolic state when a new constraint is generated and therefore maintains a mapping from constraints to their code locations. When constructing the constraint dependency tree, if two basic blocks are connected in the control flow graph, the constraint of the successor node will be dependent on that of the predecessor node. 


{\small
	\begin{lstlisting}[
	caption={An example of high-level constraints recording},
	captionpos=b
	language={C},
	label={lst:high-level},
	keywordstyle=\color{blue!70},
	numbers=left,                    
	numbersep=5pt, 
	xleftmargin=8pt,
	xrightmargin=5pt,
	numberstyle=\scriptsize\color{gray},
	breaklines=true,
	frame=single,
	basicstyle=\ttfamily, 
	commentstyle=\color{blue} \textit,
	stringstyle=\ttfamily, 
	showstringspaces=false,
	mathescape=true]
char* __nptr = (char*)websGetVar(param_1, "start_time");
char* __nptr_00 = (char*)websGetVar(param_1, "end_time");
int iVar5 = strcmp(__nptr, "0");
if (iVar5 == 0){
  iVar5 = strcmp(__nptr_00, "86400");
  if (iVar5 != 0) goto LAB_0046e540;
  else{
    //do something at 0x46e534
  }
}
	\end{lstlisting} 
}

\section{Normal User View Emulation}

For the sake of finding services that are transparent to normal users. It is essential to realize what a normal user sees at the client side. In this section, we illustrate the hybrid approach of \system to do normal user view emulation. 

Previous work~\cite{DBLP:conf/ccs/CostinZF16} has summarized several approaches of running web interfaces. Different from their scenario of penetrating the firmware by web interfaces, we only care about thoroughly collecting network requests that can be triggered by normal users, and leave the CGI-like binaries to the symbolic execution module of \system, and therefore, we propose the following strategies:

\noindent \textit{\textbf{Strategy 1.}} \textbf{Local Web Server Hosting.} We referenced the idea of how to run web interfaces in~\cite{DBLP:conf/ccs/CostinZF16}. After locating the document root in the file filtering phase, \system launches a simple web server under the document root directory using the \textit{Express} module provided by \textit{node-js}~\cite{url_nodejs}. With the web server, all \textit{html} and \textit{js} files can be accessed. \system then takes a DFS-based iteration approach to traverse all these interface files and collect generated network requests.

\noindent \textit{\textbf{Strategy 2.}} \textbf{Qemu Emulation.} Strategy 1 can only solve the issue of \textit{html} and \textit{js} file emulation. For web interface files that need the support of certain modules, like \textit{asp} and \textit{php}, we decide to take advantage of tool \textit{Firmadyne} proposed by Chen et al.~\cite{DBLP:conf/ndss/ChenWBE16}. Firmadyne provides a system-level emulation given a firmware. Using this approach, \textit{asp} and \textit{php} files can both run due to the successful recovery of its web server environment.

\noindent \textit{\textbf{Strategy 3.}} \textbf{Static Analysis.} Firmadyne has a limitation of run-time successful rate due to various layouts of firmware samples. When Firmadyne fails to emulate a firmware, we leverage the signature-based static analysis to uncover user requests. 
\section{Evaluation}
We evaluate \system on 3 firmware from two different vendors, D-Link and Netgear. Two of them have been reported as vulnerable. Our goal is to test whether we could find security-critical issues with the help of \system.

\subsection{D-Link DIR-619L wireless N300 Cloud Router}

D-Link has a long history in home-use router development and manufacturing. DIR-619L is a popular brand released by D-Link in 2012. According to statistics provided by~\cite{url_dir619_vuln}, there are in total three vulnerabilities presented by previous researchers from 2015 to 2018, whose vulnerability types are all \textit{Exec Code}.

We use the firmware of version 2.06.B01 to conduct the evaluation. After unpacking the firmware, \system first locates the document root path as \textit{/web} with 168 \textit{asp} files, 3 \textit{html} files and one \textit{cgi} file. Unfortunately, both the strategy 1 and strategy 2 fail to emulate the normal user view, \system then takes the third approach -- static analysis. After statically analyzing the web interface files, \system recovers 758 requests in total, including 662 \textit{asp} requests and 96 \textit{html} requests (html form request). In the table~\ref{tab:asp_handler}, we show the top 5 most-frequently invoked asp handlers and html urls by this firmware.

\begin{table}[htb]
\scalebox{1.0}[1.1]{
	\begin{tabular}{|c|c|c|c|}
		\hline
		\textbf{ASP handlers} & \textbf{times} & \textbf{HTML URLs}          & \textbf{times} \\ \hline
		getInfo               & 170            & /goform/formEasySetupWizard & 8              \\ \hline
		getIndexInfo          & 157            & /goform/formdumpeasysetup   & 3              \\ \hline
		firewallRule\_row     & 50             & /goform/formSetEasy\_Wizard & 2              \\ \hline
		getWizardInformation  & 35             & /goform/formWlSiteSurvey    & 2              \\ \hline
		staticRouteList       & 32             & /goform/form\_mydlink\_sign & 2              \\ \hline
	\end{tabular}
}
\vspace{0.1cm}
\caption{top-5 most-widely invoked ASP handlers and HTML URLs in DIR-619L}
\label{tab:asp_handler}
\end{table}

\system then runs its service recognition module and symbolic execution engine on the filtered service binary, which is \textit{boa} in this case. The results show that all asp handlers and HTML urls are captured by \system due to the function used to define them \textit{websAspDefine} and \textit{websFormDefine}  have been recognized as \textit{request-handler binders}. 

{\small
	\begin{lstlisting}[
	caption={The usage of \textit{websAspDefine} and \textit{websFormDefine}},
	captionpos=b
	language={C},
	label={lst:high-level},
	keywordstyle=\color{blue!70},
	numbers=left,                    
	numbersep=5pt, 
	xleftmargin=8pt,
	xrightmargin=5pt,
	numberstyle=\scriptsize\color{gray},
	breaklines=true,
	frame=single,
	basicstyle=\ttfamily, 
	commentstyle=\color{blue} \textit,
	stringstyle=\ttfamily, 
	showstringspaces=false,
	mathescape=true]
void websAspInit(void){...
 websFormDefine("formSetHNAP11", formSetHNAP11);
 websAspDefine("getInfo",getInfo);
 websAspDefine("getIndexInfo",getIndexInfo);
... }
void formSetHNAP11(char* param1){...}
void getInfo(char* param1){...}
void getIndexInfo(char* param1,char* param2,char* param3){...}
	\end{lstlisting} 
}

We then analyze the results given by \system to see if there are ``hidden'' services inside the firmware. To our surprise, \system emits two alerts:

1) Accessing ``/common/info.cgi'', ``/version.txt'' and ``DevInfo.txt'' can lead to information disclosure, some of which are sensitive, like the WPS password of the current WiFi.

After checking the function invocation trace we find that this hidden service is of the type ``persistent output'' since a call to \textit{open} with write capability(0x302) is captured by \system.

2) A function table structure is recognized at address 0x4f9014. The symbolic execution module of \system in total recovers 21 handlers from the function table. We then observe that this function table is looked up inside the form handler of url ``formSetHNAP11'', which is also not used by web interface files. Our manual code analysis reveals that such services can be utilized to set new passwords or DoS the device.

\subsection{D-Link DIR-823G Dual Band Smart Wireless Router}

The firmware we use to evaluate \system is version A1 FW102B03, \system first recognizes a request-handler binder \textit{sub\_40b1f4}. Tracking the usage of that function, we find handlers of four types of urls -- ``/HNAP1/*'', ``/goform/*'', ``/cgi-bin/*'', ``/EXCU\_SHELL''.

Different from DIR-619L, DIR-823G firmware can be successfully emulated by Firmadyne. Using the automatic triggering scripts, \system collects network requests generated by a normal user when accessing the web interface.

After comparing against the services found by \system, we find ``/EXCU\_SHELL'' is a hidden service which can be utilized to remotely execute arbitrary Linux commands. After searching recent CVE reports, we found the exploitation has been released in 2018~\cite{url_cve_goahead} but tested with different brands, which also reveals the code reuse issue among D-Link devices.

\subsection{D-Link 8010LH Wi-Fi Camera} 

8010LH is a recent Wi-Fi camera released by D-Link. Different from other two samples, \system does not find any web servers inside the firmware, however, there are three service binaries found -- ``Rtk\_MainProc'', ``StreamProxy'' and ``TW\_Monitor''. All of them are recorded inside the system initialization configuration file ``/init.d/rcS'' to launch during the starting process.

\system found one hidden service inside ``Rtk\_MainProc'' belonging to system command execution. Tracking the traces output by \system, we find that inside the ``main'' function of ``Rtk\_MainProc'', when the binary detects that file \textit{.tw\_enable\_telnet} exists, telnet service will be enabled by \textit{system(``telnetd \&'')}. We further try cracking the credentials of \textit{root} user and find the password can be cracked within two seconds.

\system finds another hidden service inside the service binary ``StreamProxy'' that creates new network activities. By tracing back from the call to ``SSL\_connect'', \system finally recovers a HTTP packet that can establish an unauthenticated RTSP channel with a remote server. The recovered HTTP request is as the listing~\ref{lst:poc}.

{\small
	\begin{lstlisting}[
	caption={Recovered HTTP request},
	captionpos=b
	language={C},
	label={lst:poc},
	keywordstyle=\color{blue!70},
	numbers=left,                    
	numbersep=5pt, 
	xleftmargin=8pt,
	xrightmargin=5pt,
	numberstyle=\scriptsize\color{gray},
	breaklines=true,
	frame=single,
	basicstyle=\ttfamily, 
	commentstyle=\color{blue} \textit,
	stringstyle=\ttfamily, 
	showstringspaces=false,
	mathescape=true]
[LAN ip]:7000/command=rtsp_start_viewing
&rtsp_host=[remote IP]
&rtsp_port=[remote port]
&code=0
	
	\end{lstlisting} 
}

\section{Related Work}

Firmware analysis has become a hot topic in recent years. Costin et al.~\cite{DBLP:conf/uss/CostinZFB14} conducted a large-scale security assessment on a dataset with more than 320,000 firmware samples. Their results show vulnerabilities are pervasive in firmware and should be attached with enough notice. Shoshitaishvili et al.~\cite{DBLP:conf/ndss/Shoshitaishvili15} proposed an automatic tool called \textit{Firmalice} to detect authentication bypass vulnerabilities in a firmware. Their work shed a light on symbolically-executing firmware codes. Furthermore, their group presented a binary code analysis framework called \textit{angr} in~\cite{DBLP:conf/csfw/BrumleyWJS07}, which is also the back-end of our framework.
D. Chen et al.~\cite{DBLP:conf/ndss/ChenWBE16} presented a qemu-based emulation tool \textit{Firmadyne}. The tool can globally simulate a firmware and well capture the behaviors of that firmware. However, it still suffers from the success rate, due to the variety in firmware layout. 
Redini et al.~\cite{redini2020karonte} proposed a taint-based static analysis tool capable of finding insecure multi-binary interactions. However, they get rid of all firmware sample of MIPS architecture, which takes up more than 79.4\% in a recent released data set~\cite{DBLP:conf/ndss/ChenWBE16}. Different from their work, our framework can well handle MIPS firmware samples.

Apart from direct analysis on firmware binaries, researchers propose other approaches targeting at different interfaces. J. Chen et al.~\cite{DBLP:conf/ndss/ChenDZZL0LSYZ18} presented a mobile-based fuzzing tool \textit{IoTFuzzer}, which can trigger vulnerabilities of IoT devices by mutating mobile-side messages. Similarly, Wang et al.~\cite{DBLP:conf/uss/WangSN019} proposed a novel approach of finding vulnerabilities inside firmwares by analyzing companion mobile apps.

\bibliographystyle{IEEEtranS}
\bibliography{ref}




%

\end{document}